# Exploring Ultra Low-Power on-Chip Clocking Using Functionality Enhanced Spin-Torque Switches

Mrigank Sharad and Kaushik Roy
Department of Electrical and Computer Engineering, Purdue University, West Lafayette, IN, USA
( msharad, kaushik)@purdue.edu

**Abstract:** Emerging spin-torque (ST) phenomena may lead to ultra-low-voltage, high-speed nano-magnetic switches. Such current-based-switches can be attractive for designing low-swing global-interconnects, like, clocking-networks and data-buses. In this work we present the basic idea of using such ST-switches for low-power on-chip clocking. For clocking-networks, Spin-Hall-Effect (SHE) can be used to produce an assist-field for fast ST-switching using global-mesh-clock with less than 100mV swing. The ST-switch acts as a compact-latch, written by ultra-low-voltage input-pulses. The data is read using a high-resistance tunnel-junction. The clock-driven SHE write-assist can be shared among large number of ST-latches, thereby reducing the load-capacitance for clock-distribution. The SHE assist can be activated by a low-swing clock (~150mV) and hence can facilitate ultra-low voltage clock-distribution. Owing to reduced clock-load and low-voltage operation, the proposed scheme can achieve 97% low-power for on-chip clocking as compared to the state of the art CMOS design. Rigorous device-circuit simulations and system-level modelling for the proposed scheme will be addressed in future.

## I. Introduction

With the scaling of CMOS technology, energy-efficiency and performance of the on-chip global-interconnects like, clocking-networks and long-distance data-links, degrade due to increase in per-unit length resistance of long metal-lines [1]. On the contrary, the increasing complexity of synchronous multi-core systems, translates to a parallel increase in the complexity and performance-needs for such long-distance links. The integration of multiple processing-cores and larger on-chip memory-blocks, for instance, has resulted in increasingly busy on-chip data-links and connection-networks for memory-access and long-distance inter-block links [1]. Hence, a major fraction (~50%) of the total power consumption in highly synchronous systems, such as chip-multi-processors (CMP), may be ascribed to such global-links [2]. As a result, the design of on-chip global interconnects has emerged as a major challenge for high-speed computing-systems and call for innovative and cross-hierarchy design-solutions.

Recently, the application of low-voltage, magneto-metallic spin-torque (ST) switches for ultra-low-energy compact and high-performance global interconnect design has been proposed [11]. Recent spin-torque (ST) experiments have demonstrated high-speed spin-torque switching based on Spin Hall Effect (SHE) [11-13]. Such emerging spintronic-phenomena may be conducive to the design of ultra-low-voltage, low-current and high-speed nano-magnetic switches that can be applied to the design of energy-efficient global interconnects [11].

In this work we propose a novel spin-torque switch (STS), based on SHE-effects, that can be suitable for the design of ultra-low power global-interconnects like clocking-network and data-buses. The proposed ST-switch offers conditional write, triggered by an ultra-low voltage signal. Hence it can act as a non-volatile magneto-metallic latch, driven by an ultra-low voltage, low-power clock. The simulation and modeling-based results obtained for the proposed ST-device for global-clocking show the possibility of large power savings over state of the CMOS design-solutions.

Rest of the paper is organized as follows. Section-II presents the basic ST-device proposed in this work. Application of the proposed ST-device in clocking is discussed in section-III. Section-IV describes the design of global data-link. Conclusions are given in section-V.

## II. Spin-Torque Switch for Global Interconnect

*A. Application SOC for high-speed ST switching:* Recently, several high-speed spin-torque (ST) switching phenomena have been proposed and demonstrated. High-speed magnetization switching in domain-wall magnets (DWM) [11-15] as well as spin-valves (SV) [16, 17], have been shown. In this work, we propose a high-speed spin-torque (ST) switch using Spin-Hall Effect (SHE) in a DWM-based device [11].

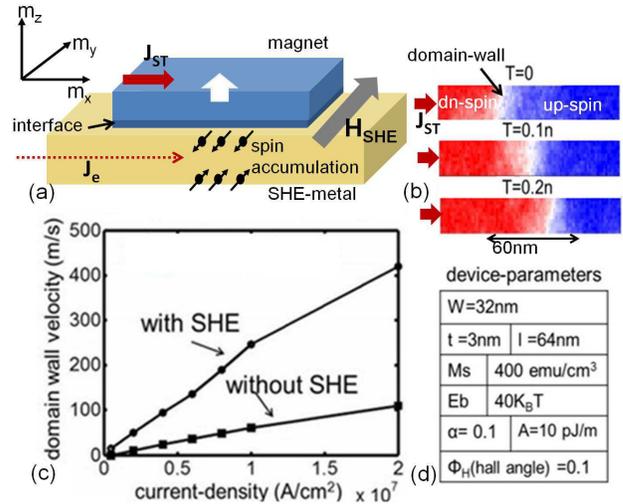

Fig.1 (a) Depiction of SHE acting on a nano-magnet layer, placed on SHM, (b) transient micro-magnetic simulation plots for the SHE assisted switching of the free-layer (Je = Jst = $3 \times 10^6 \text{A/cm}^2$), (c) improvement in domain-wall switching-speed due to SHE, (d) magnet dimensions and parameters used for results in c, d.

A magnet has two anti-parallel, stable spin-polarization states, that lie along the 'easy-axis' of the magnet [18]. The magnet shown in fig. 1a, has its easy-axis along the vertical-direction (z axis). Hence, the two stable spin-states are 'up-spin' and 'down-spin'. It is well known that the spin-polarity of a nano-scale magnet can be flipped between its two stable-states (up and down-spin in this case), by injecting spin-polarized electrons, with polarization parallel to one of the two stable-states [18]. This effect is knows as spin-torque (ST) switching. The direction orthogonal to the easy-axis, is an unstable state for the magnet's spin-polarity and is commonly termed as the 'hard-axis' (in-plane direction in fig. 1a). However, a magnetic-field applied along the hard-axis can significantly accelerate the ST-switching mechanism, by effectively lowering the energy barrier for the transition between the two stable-states [16, 17].

The principle of SHE assisted ST-switching is depicted in fig.1. The magnet in fig.1a lies on the top of a Spin-Hall Metal (SHM) [13]. A charge-current passing through an SHM layer results in accumulation of electrons of opposite spin-polarity (in-plane and out of plane in fig. 1a) along its top and bottom surfaces. This phenomenon roots from structural inversion asymmetry (SIA) in some specific heavy-metals (and alloys, termed as SHM) and is called Spin-Hall-Effect (SHE). The spin-accumulation resulting from SHE results in an effective magnetic-field ($H_{SHE}$) pointing parallel to the SHM surface and orthogonal to the direction of current-flow. The magnitude of this field can be expressed as $\alpha J_e P/\mu_B M_s$ [13], where, where $\mu_B$ is the Bohr-magneton and $P$ is the polarization of the carriers at the SHM surface, ferromagnetic layer and α is constant determining the effectiveness of SHE (discussion on optimum values of these parameter values can be found in [13]). The SHE-field, being parallel to the hard-axis of the magnet, can assist current-induced ST-switching. For instance, a spin-polarized electron-current $J_{ST}$ flowing along the magnet, as shown in fig. 1a, can cause fast ST-switching of the magnet, in the presence of $H_{SHE}$. The ST-switching of a magnetic layer due to an in-plane current (like $J_{ST}$ in fig.1a), proceeds through the motion of magnetic domain-wall (DW) along the magnet (fig. 1b). A DW is effectively a transition-region between magnetic-domains of opposite-polarity. The SHE-assist can improve the ST-switching speed of DW, for a given ST-current, by more than an order of magnitude [14]. Possibility of ~10ps switching speed for magnets has been predicted with SHE assist [19].

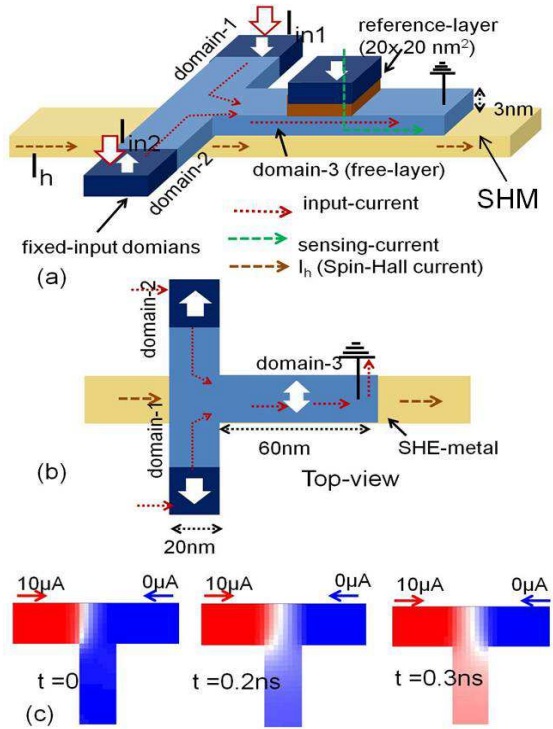

Fig. 2 (a) STS on SHE-assist (b) top-view of the device, (c) micromagnetic simulation plots for the STS at three-time steps
Next, we discuss the proposed ST-switch based on SHE-assist.

B. *High-Speed ST-Switch (STS) based on SOC*

A 3-terminal, spin-torque switch (STS) based on SHE is shown in fig. 2a. It consists of two fixed-domains of opposite magnetization (domain-1 and domain-2) that act as input-ports and spin-polarize the input currents. The third-domain (domain-3) is a free-domain. The overall spin-polarity of the current injected into the free-domain is parallel to the input-domain which receives larger-current. The free-domain can switch parallel to either of the two fixed input-domains, depending upon which of the two inputs-currents is larger. Hence, the STS-device acts as a compact and ultra-low voltage current-comparator. The minimum current-required to switch the STS depends upon the critical current-density for domain-wall (DW) shift in the free-domain. Under the influence of SHE assist field, ~100m/s DW-velocity can be achieved with a current-density of ~$10^6$ A/cm$^2$ [13]. This implies that, a 100nm long free-domain with cross-section-area 22x3nm$^2$ can be switched within ~1ns with a current of the order of few micro-amperes. Larger currents can lead to faster-switching. To incorporate the effect of SHE-assist in the proposed device, the free-domain of the STS is placed on the top of an SHM strip (fig. 2a), carrying an equivalent current-density. This assist-current can be injected into the SHE-layer through a separate input terminal associated with the SHM-strip.

Due to significant contact resistance between the SHM and the magnet, the current flowing along the two materials may be controlled independently. It has been shown that, $H_{SHE}$ can be equally effective even when the SHM and the magnet are isolated by a thin-spacer [14]. In this work, we have assumed the current-flow in the free-domain and the SHM can are independent. This can be facilitated by including a resistive spacer between the two layers. Micro-magnetic simulation results for an input of 10μA (the other input being zero) is shown in fig. 2c.

The state of free-domain (domain-3) is 'read' through the magnetic tunnel junction (MTJ) formed at its top (with a fixed-reference magnet-layer, as shown in fig. 1a) [11]. The resistance of an MTJ is high when the two magnetic-layers possess anti-parallel spin-polarity and vice-versa. MTJ-resistance-ratios of greater than 6 have been reported in literature [20].

Notably, the proposed device offers a low-resistance magneto-metallic path to the switching-current and hence, can operate with very small input-voltages. The high-resistance MTJ port can provide large-output-voltage levels (high or low, depending upon anti-parallel or parallel state of the MTJ) using a small sensing-current. In the following sections we describe the application of the proposed STS device in the design of energy-efficient global-interconnects like, clocking-network and data-buses.

### III. Ultra Low-Power Clocking Scheme Using Spin-Torque Switch

In this section we present the design of low power latch using STS and discuss its circuit and system-level benefits for energy-efficient clocking.

*A. Circuit for STS-latch and write-operation:* Having a separate input-port for the SHE-assist, the STS device offers the possibility of conditional switching, for a given set of input currents. By stopping the current-flow in the SHM layer, the intrinsic switching threshold of the STS device can be enhanced. Hence, it can fail to switch for input currents below this intrinsic threshold (fig. 3). This phenomena can be exploited to model a compact, ultra-low-voltage and non-volatile magnetic-latch using the STS, as shown in fig4.

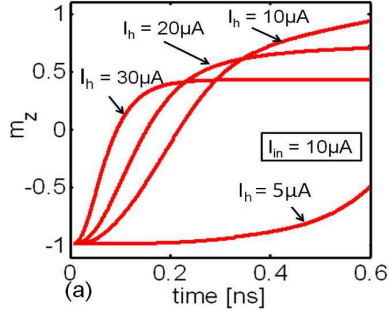

Fig. 3 Switching of STS free-layer (domain-3)(for 0.5ns duration) under different magnitudes of SHM current $I_h$, showing hindered switching for lower $I_h$. Too high $I_h$ for a given inputs forces the magnetization away from the easy-axis ($m_z = +/-1$).

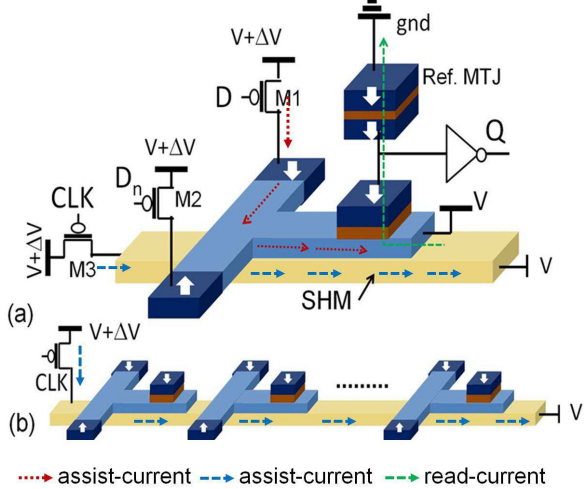

Fig. 4 (a) circuit for non-volatile magnetic latch using STS, (b) multiple STS latches sharing the SHM-strip to reduce clocking power (both static and dynamic).

In the proposed latch, the SHE input acts as the clocking signal and is supplied through the transistor M3 in fig.4a. The input-data and its complement connect to the two input domains, through transistors M1 and M2. The STS-free-layer switches in presence of the SHE-clock, depending upon the value of the input bit $D_n$. Thus, this circuit acts as a level-sensitive latch. Owing to the low-resistance switching-current-path (both, magnet and the SHM) in the STS, the input transistors (and the clocking transistor) can be biased at a very small drain-to-source voltage (ΔV < 100mV) [11]. This can help achieve low-power current-mode write-operation for the STS-latch.

*B. Read-operation for STS-latch:* The state of the latch is read using a resistive voltage-divider formed between the STS-MTJ and a fixed reference-MTJ (fig. 4a). The resistance ratio of an MTJ is defined in terms of tunnel magneto-resistance ratio (TMR) as: $(R_{AP}-R_P)/R_P \times 100$, where $R_{AP}$ and $R_P$ are the anti-parallel and the parallel-state resistances of the MTJ respectively. For a TMR of ~200%, a voltage swing of ~V/3 can obtained using the voltage divider (where V is the supply-voltage). Such an output swing can be directly detected by a simple CMOS inverter. Higher TMR may provide higher output-swing and hence better robustness (fig. 5a). High oxide thickness ($t_{ox}$) for the MTJ provides higher absolute resistance for the voltage-divider, minimizing the leakage-power in the read-operation. However, too high value for MTJ resistance diminishes the output-swing for high-frequency operation, due to low-pass filtering effect (fig.5b).

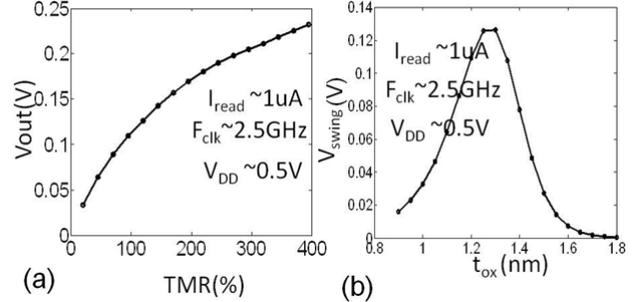

Fig. 5(a) increase in output swing with TMR, (b) Effect of $t_{ox}$ on MTJ output swing

*C. Energy-dissipation of a single STS-latch* Simulations (using 45nm CMOS) show that the STS-static-power (including the inverter) can be limited to a fraction of a micro-Watt for the STS-latch (during active-mode) for 2.5 GHz operation, implying less than 0.4fJ energy-dissipation. The energy-dissipation due to the write operation can be determined based on the write-current and ΔV. A ~10µA current through ~100mV consumes ~0.4fJ as well. The total energy-dissipation for the STS-latch was more than ~10x lower as compared to state of the art CMOS latch (~15fJ in 45nm CMOS).

*D. Sharing SHM assist for low clocking power:* The clock signal driving the SHE current can be shared between multiple SHE-latches, as shown in fig. 4b, by sharing the SHM strip among the latches. This reduces the switched-capacitance power for the driving clock (represented by the clock driven transistor M3 in fig.4a), along with the static-clock-power (due to SHM-current). However, the effective resistance of the SHM strip would increase linearly with the number of latches sharing an SHM-strip. This may require increase in ΔV. However, SHM materials like Cu-Bi offer relatively low-resistivity [21], and can facilitate sharing of SHM strip among more than ~20 STS latches (~1µm long Cu-Bi-based SHM-strip, shared by ~20 latches was estimated to offer ~3kΩ of resistance, which is small enough to provide ~20µA of SHE-current with ΔV of ~100mV), without significant increase in the required ΔV.

*E. Optimizing clocking transistor and clock voltage:*
Clocking-power can be further reduced by optimizing the clocked-transistor in the STS latch, while considering the specific characteristics of ST-switching. Magnets offer an inherent switching-threshold for a given switching-time. The same is true with respect to the SHE assist. A factor of 5-6 reduction in the SHE current was found to suffice for preventing switching in STS under room-temperature simulations (notably read-operations in spin-torque memory, generally employs a read-current close of 0.33x that of ST-write-current for similar read-write time). Thus, the clocking transistor can afford to have a poor on-off ratio of ~5. This fact can be exploited by employing a small voltage-swing for the clock signal, enough to produce a current-swing of ~5x in the clocked transistor. The $V_T$ of the transistor can be appropriately adjusting according to the choice of the bias condition (for example, a source/drain doping can be used to lower the $V_T$; drastic reduction in $V_T$ is expected to degrade the on-off current-ratio, but such a degradation can be tolerated by the ST-device), such that it can provide a current-swing of ~5x for a relatively small gate-voltage swing. In this work, we optimized the driving transistors for a gate drive of ~100mV (same as $\Delta V$). The use of low-voltage clock along combined with, reduced-load capacitance (due to sharing of SHM strip), can therefore bring huge reductions in overall clocking power. The system-level rationale of this approach is discussed in the following sub-section.

*F. System-Considerations:*
Although, edge-triggered flip-flops (FF) can be designed using the proposed STS-latch, in this work we explore level-sensitive pulsed clocking scheme that employs latches rather than FFs (fig. 6).

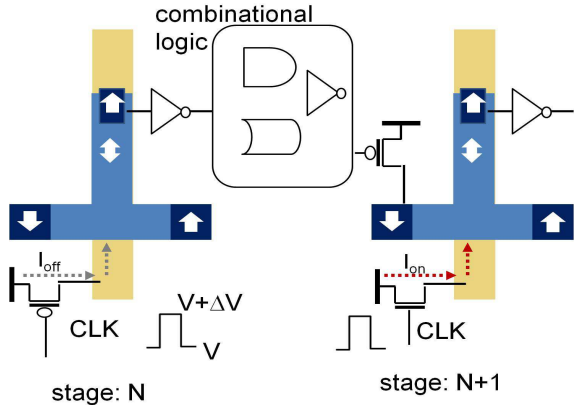

Fig. 6 (a) Pulsed clocking scheme using STS: uses only one-clock phase.

Latch-based pulsed clocking offeres several advantages over FF-based clocking (like, higher performance, time borrowing etc) [22]. In this work we employed PMOS and NMOS transistors to drive alternate stages of an STS based pipeline (fig.6). The threshold voltages for both the driving transistors are adjusted to provide an on-off current ratio of ~5 with ~100mV swing, at opposite clock-phases. For better robustness, two non-overlapping clock phases can also be used.

Fig. 7 depcits the schematic for one of the popular state of the art clocking-system, that employs a combination of H-tree and global clock-mesh [3]. The H-tree drives global clock-buffers which in turn flood the global clock-mesh with the clock-signal from numerous locations. The different clock-meshes on a chip may be shorted to achieve better de-skewing across the processor. On-chip inductors have been used to resonate with the mesh capacitance power, in-order to reduce the global-clock power [3]. Local clock buffers (LCB) receive the mesh-clock signal and deliver it to the local-loads, constituted by the flip-flops (/latches). It has been reported that the effective load-capacitance faced by the LCBs is generally more than ~10x higher than that of the clock-mesh itself [23]. Hence, the LCBs and the latches account for most of the clock-power.

In section-III-D, we noted that the STS-latch can help reduce the effective load capacitance by more than an order of magnitude. This may result in the total latch capacitance close to or even significantly smaller than that of the global-clock-mesh. Thus, it might be possible to drive the STS-latches directly using an ultra-low-swing global-clock, thereby eliminating the overhead due to the LCBs. An important consideration in this case, however, may be the shape of the clock-signal received at the STS latches. Since, the pulsed-clocking scheme is sensitive to level (and not so much to the clock-edge declivity), it is generally enough to ensure sufficient non-overlap between the two clock-phases [22]. A simple possible circuit for mesh-driver is shown in fig. 8a. It employs linear region-PMOS transistors to supply the $\Delta V$-clock to the mesh (note: the characteristics of such a driver is completely different from an inverter, which would fail to operate with such a low-terminal voltage and large capacitive load).

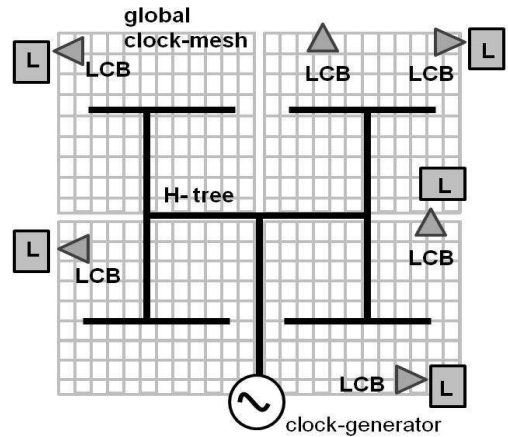

Fig. 7 State of the art clocking system based on a combination of H-tree and global clock-mesh

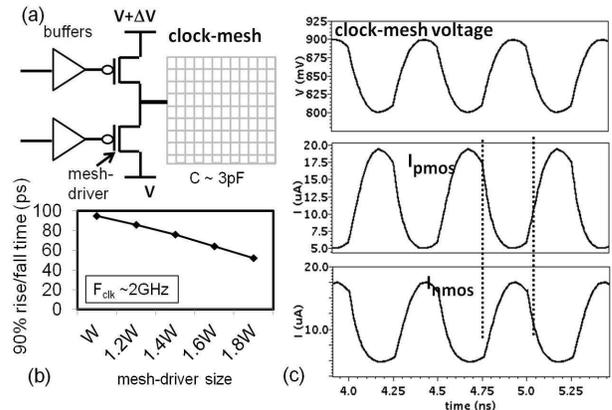

Fig. 8 (a) global clock-mesh and associated driver for ultra low-voltage STS-based clocking scheme, (b) effect of mesh-driver size of the rise/fall time of mesh clocks directly received by the STS-latches, (transient waveform for mesh-clock and clocked-current inputs to two consecutive SHL-latches

*G. Overall Benefits of STS-based clocking:* The foregoing discussion on system-level implications of the proposed clocking scheme indicates the possibility of large energy-saving for the entire clocking-system, including the latches. STS-latches can offer reduced effective load-capacitance (which can be now almost same as the total mesh-capacitance) and can also facilitate the use of ultra-low voltage clock-swing. As a result, the power of the H-tree and the buffers driving the mesh-drivers are left as the dominant power-consuming circuits in the entire clock-network. Their power consumption was found to be less than 3% of the total clock-tree power. Thereby implying more than 97% power saving [23].

Another important advantage of the proposed design is the non-volatility of the STS-latches. At scaled CMOS technology nodes ~50% of the latch power can be ascribed to the idle-leakage power. Several design solutions have been proposed to reduce this component (like, the use of high-$V_T$ CMOS latches for preserving data in 'sleep-mode' [24]). However, due to its in-built non-volatility, the STS latch can be gated more efficiently.

In the next section we discuss the application of STS in the design of the other very important class of global on-chip-links, namely global data-interconnects.

## V. Conclusion

Recent experiments have shown the direction for ultra-high-speed spin-torque switching. Such phenomena can be exploited in modeling compact, ultra-low-voltage current-mode switches. ST-based current-mode switches can facilitate the design ultra low power and high-performance global links, like, clocking networks and global-data-interconnects. In this work we proposed a novel ST-device and analyzed its applications in global-clocking. We employed physics based micro-magnetic simulation-model for the proposed-device [26]. A behavioral circuit-model was used for SPICE simulations. Results showed the possibility of ~97% improvement in energy for on-chip clocking can be achieved with optimal device parameters.